\documentclass[12pt]{iopart}
\usepackage{graphicx,iopams}
\usepackage{citesort}
\usepackage{color}
\usepackage{lineno}

\begin{document}


\title{Instability dynamics of nonlinear normal modes in the Fermi-Pasta-Ulam-Tsingou chains}
\author{Liangtao Peng$^1$,~Weicheng Fu$^{2,3,\dag}$,~Yong Zhang$^{1,3,\ddag}$,~Hong Zhao$^{1,3}$}
\address{$^1$Department of Physics, Xiamen University, Xiamen 361005, Fujian, China\\
$^2$Department of Physics, Tianshui Normal University, Tianshui 741001, Gansu, China\\
$^3$Lanzhou Center for Theoretical Physics, Key Laboratory of Theoretical Physics of Gansu Province, Lanzhou University, Lanzhou 730000, Gansu, China}
\ead{$^\dag$fuweicheng@tsnu.edu.cn;~$^\ddag$yzhang75@xmu.edu.cn}


\begin{abstract}
Nonlinear normal modes are periodic orbits that survive in nonlinear chains, whose instability plays a crucial role in the dynamics of many-body Hamiltonian systems toward thermalization. Here we focus on how the stability of nonlinear modes depends on the perturbation strength and the system size to observe whether they have the same behavior in different models. To this end, as illustrating examples, the instability dynamics of the ${N}/{2}$ mode in both the Fermi-Pasta-Ulam-Tsingou (FPUT) -$\alpha$ and -$\beta$ chains under fixed boundary conditions are studied systematically. Applying the Floquet theory, we show that for both models the stability time $T$ as a function of the perturbation strength $\lambda$ follows the same behavior; i.e., $T\propto(\lambda-\lambda_c)^{-\frac{1}{2}}$, where $\lambda_c$ is the instability threshold. The dependence of $\lambda_c$ on $N$ is also obtained. The results of $T$ and $\lambda_c$ agree well with those obtained by the direct molecular dynamics simulations. Finally, the effect of instability dynamics on the thermalization properties of a system is briefly discussed.
\end{abstract}

\section{\label{sec:1}Introduction}

Periodic orbits play an important role in many-body Hamiltonian systems with nonlinear interactions. These orbits can lead to a long-time metastable state, which have been one of the core concerns in statistical mechanics \cite{2008LNP728G,bookComplex2012}. A well-known example is the Fermi-Pasta-Ulam-Tsingou (FPUT) recurrence phenomenon found in the 1950s \cite{fermi1955studies}, which showed that a one-dimensional(1D) anharmonic chain far from equilibrium does not enter the expected thermalized state over time, but instead returns to nearly the initial state of nonequilibrium. This recursive phenomenon is generally attributed to the time-periodic localized structures in normal mode space called $q$-breathers \cite{PhysRevLett.95.064102,PhysRevE.73.036618}, or more generally $q$-tori \cite{PhysRevE.81.016210}, which are exact solutions of the system studied.

Nonlinear normal modes (NNMs) are another kind of simple periodic structures in 1D anharmonic chains, which correspond to stationary waves of the form $x_j(t)=R_k(t)Q_{j,k}$, where $R_k(t) $ is a nonlinear oscillatory function of time \cite{poggi1997exact}, and the components $Q_{j,k}$ is known as the linear normal modes. Therefore, the displacement $x_j(t)$ of every particle is proportional to the displacement of an arbitrary chosen particle \cite{chechin2012stability}. In general, these NNMs are mainly found through the following two methods. The one is by analyzing the selection rules derived from the equations of motion in the Fourier space \cite{poggi1997exact,BIVINS197365}, and the other is to analyze the discrete symmetry of the chain model by using the group theory \cite{CHECHIN2002208,CHECHIN2005121,RINK200331,chechin2012stability}.

The stability problem of NNMs has been widely studied \cite{PhysRevE.70.016611,PhysRevE.74.047201,leo2007stability,PhysRevE.94.042209,poggi1997exact,PhysD1983,BIVINS197365,chechin2012stability,CHECHIN2002208,CHECHIN2005121,RINK200331,PhysRevE.69.046604,PhysRevE.73.056206}.
The stability of the NNMs can be detected numerically through integrating the Newtonian dynamics of the whole system \cite{PhysRevE.69.046604} or studied theoretically by analyzing the variational dynamics of the NNMs by the Floquet theory \cite{PhysD1983}. As a result, the instability threshold as a function of system size $N$ is obtained. Especially in the large $N$ limit, it has been reported that the instability threshold decreases in a power-law manner with the increase of $N$; i.e., $N^{-\zeta}$, and $\zeta=1$, or $2$ is decided by both the boundary conditions and the wave number of the NNMs \cite{chechin2012stability}. It has been suggested that there is a certain relationship between the instability of specific NNMs and the emergence of global chaos in the system \cite{PhysRevE.73.056206}. For instance, if $\zeta=1$ for a specific mode, then its destabilization is related to the transition from weak to strong chaos. On the other hand, if $\zeta=2$ for a specific mode, it is connected with the onset of weak chaos as a result of the breakdown of the FPUT recurrences (detailed discussions see \cite{bookComplex2012}). Whereas it also has been proposed that the destabilization of the NNMs leads to the formation of stochastic layers, but their thickness and extension in phase space remain small at weak perturbations \cite{poggi1997exact}. Obviously, the dependence of the instability threshold on $N$ is not enough to determine the role of the NNMs in the structure of the phase space of nonlinear many-body Hamiltonian systems, which can not reflect the details of diffusion of NNMs in phase space. Thus, the instability dynamics of these NNMs need to be further studied in detail.

It is noted that the dynamics of NNMs are decoupled from all other modes under appropriate initial conditions, that is, if only one of them is initially excited, its evolution will not transfer energy to any other modes, which means that the system will not reach the thermalized state. Although some recent research shows that the nontrivial multi-wave interactions can lead to the FPUT chains eventually entering the state of thermalization \cite{Onorato4208,PhysRevLett.120.144301,EPL2018Thermalization}, especially more recent studies show that a general Hamiltonian system in the near-integrable region exhibits a universal thermalization behavior in the thermodynamic limit, that is, the thermalization time of the system is inversely proportional to the square of the perturbation strength \cite{fu2019universal,Fu2019PRER,Pistone2018,PhysRevE.100.052102,PhysRevLett.124.186401,2020arXiv200503478W,PhysRevE.104.L032104,Feng_2022}. However, it is found via careful investigation of these studies that the initial conditions adopted in these work are to excite multiple modes, namely, the initial excitation conditions meet the preconditions of the theoretical analysis approach of multi-wave interactions. If these NNMs are initially excited, the prerequisite for the system to enter the thermalized state is that these NNMs lose stability first so that other modes obtain energy, and then the system enters the thermalized state through multi-wave interactions. Therefore, in order to understand the nature of the system tending to equilibrium, it is necessary to carefully study the instability dynamics of these NNMs.

In the present work, we revisit the instability dynamics of the NNMs, focusing on how the stability time depends on the perturbation strength, and exploring whether they have the same behavior in different models. To this end, we take the famous Fermi-Pasta-Ulam-Tsingou (FPUT) -$\alpha$ and -$\beta$ models as examples to systematically study the instability dynamics of the ${N}/{2}$ mode which is jointly owned by these two models under fixed boundary conditions. Comparing the instability dynamics of the NNMs in these two models is mainly motivated by the fact that the FPUT-$\alpha$ model is very different from the FPUT-$\beta$ model in many aspects, such as the thermal expansion property \cite{kittel2004introduction}, the process of energy equipartition \cite{Fu2019PRER}, and the behavior of energy transport \cite{lepri2016thermal}. In the following sections, we first introduce the models and the $N/2$ mode solutions in the Sec.~\ref{sec:2}, then conduct the theoretical study on stability of the $N/2$ mode by applying the Floquet theory in Sec.~\ref{sec:3}. Our numerical results of molecular dynamics simulations (MDS) are provided in Sec.~\ref{sec:4}, followed by the conclusions and discussions in Sec.~\ref{sec:5}.

\section{\label{sec:2}The models and the $N/2$ mode solutions}

\subsection{The models}

We consider a nonlinear lattice consisting of $N$ particles of unit mass with nearest-neighbor interaction, whose Hamiltonian can be generally written as
\begin{equation}\label{eqHam}
  H=\sum_{j=1}^{N}\left[\frac{p_{j}^{2}}{2}+\frac{\left(x_{j}-x_{j-1}\right)^{2}}{2}+\frac{\theta_{n}}{n}\left(x_{j}-x_{j-1}\right)^{n}\right],
\end{equation}
where $p_j$ and $x_j$ are, respectively, the momentum and the displacement from the equilibrium position of the $j$th particle, $\theta_{n}$ denotes the nonlinear coupling strength, and $n$ is the order of the anharmonic interaction. Hereafter, we only focus on the two cases: $n=3$ for the FPUT-$\alpha$ model ($\theta_3\mapsto\alpha$); and $n=4$ for the FPUT-$\beta$ model ($\theta_4\mapsto\beta$). Based on different considerations, Refs. \cite{PhysRevE.69.046604} and \cite{fu2019universal} consistently show that the dimensionless perturbation strength of the FPUT-$\alpha$ model is
\begin{equation}
  \lambda = \alpha^2\varepsilon,
\end{equation}
where $\varepsilon$ is the energy density, i.e., the energy per particle. For the FPUT-$\beta$ chain, the dimensionless perturbation strength is
\begin{equation}
  \lambda = \beta\varepsilon.
\end{equation}

From the Hamiltonian (\ref{eqHam}), we can derive the equations of motion as
\begin{equation}\label{eqMotion}
\ddot{x}_{j}=\left(x_{j+1}-x_{j}\right)-\left(x_{j}-x_{j-1}\right)
+\theta_{n}\left[\left(x_{j+1}-x_{j}\right)^{n-1}-\left(x_{j}-x_{j-1}\right)^{n-1}\right].
\end{equation}
For convenience of discussion, we introduce the normal
mode through
\begin{equation}\label{3b}
x_{j}(t)=\sqrt{\frac{2}{N}} \sum_{k=1}^{N-1} Q_{k}(t) \sin \left(\frac{jk\pi}{N}\right),
\end{equation}
for the fixed boundary conditions, i.e., $x_0=p_0=x_{N}=p_{N}=0$ (there are $N-1$ moving particles), where $Q_{k}$ is the amplitude of the $k$th normal mode. Inserting Eq.~(\ref{3b}) into Eq.~(\ref{eqHam}), the Hamiltonian can be rewritten in the normal modes coordinate (see Appendix A for details), then the equations of motion become
\begin{equation}\label{6}
\ddot{Q}_{k}=-\omega_{k}^{2} Q_{k}-\frac{\theta_{n}}{(2 N)^{n / 2-1}}
\sum_{k_{2}, \cdots, k_{n}} \omega_{k} \omega_{k_{2}} \cdots \omega_{k_{n}} C_{k, k_{2}, \cdots, k_{n}} Q_{k_{2}} \cdots Q_{k_{n}},
\end{equation}
where
\begin{equation}\label{5}
\omega_{k}=2 \sin \left(\frac{\pi k}{2 N}\right),\quad 1\le k\le N-1,
\end{equation}
is the frequency of the $k$th normal mode, and
\begin{equation}\label{eq_Ckks}
  C_{k, k_{2}, \cdots, k_{n}} =
  \sum_{m=0}^{[n/2]}(-1)^{m}\sum_{\pm}\delta_{ k_1\pm k_2\cdots\pm k_n,\pm2N\cdot m}
\end{equation}
represents the selection rules of the interactions among the normal modes, where $\delta$ is the Kronecker delta function, and $[\cdot]$ means round down (see Appendix A for details).

To each mode one can associate a harmonic energy
\begin{equation}\label{4}
E_{k}(t)=\frac{1}{2}\left[P_{k}^{2}(t)+\omega_{k}^{2} Q_{k}^{2}(t)\right],
\end{equation}
where $P_k=\partial H/\partial\dot{Q}_k$ is the canonical conjugate momentum of the $k$th mode. The amplitude $Q_k$, conjugate momentum $P_k$, and energy $E_k$ of the $k$th mode satisfy the following relationship
\begin{equation}\label{36}
Q_{k}(t)=\sqrt{2 E_{k}(t) / \omega_{k}^{2}} \cos \varphi_k,\quad P_{k}(t)=\sqrt{2 E_{k}(t)} \sin \varphi_k,
\end{equation}
where $\varphi_k$ is the phase of the mode.

It is seen that, if $\theta_n=0$, the system is integrable: all normal modes oscillate independently and $E_k$ is constant for each $k$. The normal modes are instead coupled when $\theta_n \neq 0$.

\subsection{The $N/2$ mode solution for the FPUT-$\beta$ chains}

For the FPUT-$\beta$ chain, if only the $N/2$ mode (here $N$ should be an even number) was initially excited; i.e., $Q_{k}(0)=0$ for $k\neq{N}/{2}$, from Eq.~(\ref{6}) we obtain
\begin{equation}\label{11}
\ddot{Q}_{\frac{N}{2}}=-\omega_{\frac{N}{2}}^{2} Q_{\frac{N}{2}}-\frac{\beta}{N} \omega_{\frac{N}{2}}^{4} Q_{\frac{N}{2}}^{3}.
\end{equation}
It is well known that the solution of Eq.~(\ref{11}) can be written, with the Jacobi elliptic cosine function; i.e., $\rm{cn}$, in the form \cite{PhysRevE.54.5766}
\begin{equation}\label{12}
Q_{\frac{N}{2}}=\Lambda \rm{cn}\left(\Omega t, \Gamma^2\right),
\end{equation}
which has the period $T_{\rm cn}=4K(\Gamma^2)/\Omega$, and $K(\Gamma^2)$ is the complete elliptic integral of the first kind, where
\begin{equation}\label{13}
\Lambda=\sqrt{\frac{N\Gamma^2}{\beta\left(1-2 \Gamma^2\right)}},
\quad\Omega=\sqrt{\frac{2}{1-2 \Gamma^2}},
\end{equation}
and
\begin{equation}\label{eq_Gamma}
\Gamma^2=\frac{1}{2}-\frac{1}{2\sqrt{4\lambda(N-1)/N+1}},
\end{equation}
is the modulus of the Jacobi elliptic function.
The energy of the $N/2$ mode is
\begin{equation}\label{eq_En_h_beta}
E_{\frac{N}{2}}=\frac{1}{2} \left(P_{\frac{N}{2}}^2+2Q_{\frac{N}{2}}^2\right)+\frac{\beta}{N}Q_{\frac{N}{2}}^4.
\end{equation}

\subsection{The $N/2$ mode solution for the FPUT-$\alpha$ chains}
Similarly, for the FPUT-$\alpha$ chain, if only the $N/2$ mode was initially excited, from Eq.~(\ref{6}) we obtain
\begin{equation}\label{9}
\ddot{Q}_{\frac{N}{2}}=-\omega_{\frac{N}{2}}^{2} Q_{\frac{N}{2}},
\end{equation}
whose solution is
\begin{equation}\label{10}
Q_{\frac{N}{2}}=\sqrt{\varepsilon(N-1)} \cos \left(\omega_{\frac{N}{2}} t+\varphi_{\frac{N}{2}}\right),
\end{equation}
where $\varphi_\frac{N}{2}$ is the initial phase. The energy of the $N/2$ mode is
\begin{equation}\label{eq_En_h_alpha}
E_{\frac{N}{2}}=\frac{1}{2} \left(P_{\frac{N}{2}}^2+2Q_{\frac{N}{2}}^2\right).
\end{equation}

\section{\label{sec:3}The Floquet theory for stability of the $N/2$ mode solutions}

In principle, from mathematical point of view, there should not occur any energy exchange among different modes when only the $N/2$ mode is exactly excited at the initial. However, one-mode solution has intrinsic instability, so it may lose stability in the presence of perturbation. In practice, e.g., in numerical simulations, due to the computer's cut-off error or the limited accuracy of the integration algorithm, errors acting as a perturbation will inevitably arise in the process of evolution of a system. Then errors evolve with the system. In general, the evolution of these errors will show different behavior under different system parameters. For example, for some specific system parameters, the error will approach zero or always remain a finite value as the system evolves, thus the system is stable; on the contrary, if the error evolves to infinity, the system is unstable.

Assuming that only the $N/2$ mode is initially excited, from Eq.~(\ref{6}), we can get the evolution equation of the error (denoted as $\Delta Q_{k}$) on the mode $Q_{k}$ as follows
\begin{equation}\label{17}
\Delta \ddot{Q}_{k}=-\omega_{k}^{2} \Delta Q_{k}-\frac{(n-1) \theta_{n} \omega_{k}}{(2 N)^{n / 2-1}} \omega_{\frac{N}{2}}^{n-2} Q_{\frac{N}{2}}^{n-2}
 \sum_{k_{n}} \omega_{k_{n}} C_{k, \frac{N}{2}, \cdots, \frac{N}{2} ,k_{n}} \Delta Q_{k_{n}}.
\end{equation}
To demonstrate our research idea clearly, we will first take the FPUT-$\beta$ model as an example to carry out the corresponding theoretical analysis since the above equation in the FPUT-$\beta$ model is relatively simple.

\subsection{Analysis for the FPUT-$\beta$ chains}

For the FPUT-$\beta$ chain, Eq.~(\ref{17}) can be formulated as
\begin{equation}\label{37}
\Delta \ddot{Q}_{k}=-\omega_{k}^{2} \Delta Q_{k}-\frac{3 \beta \omega_{k}}{2 N} \omega_{\frac{N}{2}}^{2} Q_{\frac{N}{2}}^{2} \sum_{k_{4}} \omega_{k_{4}} C_{k, \frac{N}{2} , \frac{N}{2}, k_{4}} \Delta Q_{k_{4}},
\end{equation}
where the coefficient
\begin{eqnarray}
  C_{k, \frac{N}{2},\frac{N}{2}, k_{4}} &=&
  \delta_{k - \frac{N}{2}+\frac{N}{2} - k_{4}, 0}+
  \delta_{k + \frac{N}{2}-\frac{N}{2} - k_{4}, 0}\nonumber\\
  &~&+\delta_{k - \frac{N}{2}-\frac{N}{2} + k_{4}, 0}
  -
  \delta_{k + \frac{N}{2}+\frac{N}{2} + k_{4}, 2N}
\end{eqnarray}
is non-vanishing only when $k_4=k$ (see the first two terms) or $k_4=N-k$ (see the last two terms, one plus one minus cancel out, so they do not show up in the equation of motion). Thus, Eq.~(\ref{37}) reduces to
\begin{equation}\label{38}
\Delta \ddot{Q}_{k}=-\omega_{k}^{2}\left(1+\frac{6 \beta}{N} Q_{\frac{N}{2}}^{2}\right) \Delta Q_{k},
\end{equation}
which shows that all the modes are decoupled, where $Q_{\frac{N}{2}}$ is given by Eq.~(\ref{12}). Letting $W_1=\Delta Q_{k}$ and $W_2=\dot{W}_1$, then Eq.~(\ref{38}) can be rewritten as
\begin{equation}\label{eq_W}
\frac{d}{dt}
\left[
\begin{array}{c}
{W_1}\\
{W_2}
\end{array}\right]=
\left[
\begin{array}{cc}
{0} & {1} \\
{c(t)} & {0}
\end{array}
\right]
\left[
\begin{array}{c}
{W_1}\\
{W_2}
\end{array}\right],~\rm{or}\quad \frac{d}{dt} \boldsymbol{W}=\boldsymbol{A}(t)\boldsymbol{W},
\end{equation}
where
\begin{equation}\label{eq_Cbeta}
  c(t)=-\omega_{k}^{2}\left[1+\frac{6 \Gamma^2}{1-2\Gamma^2} \rm{cn}^2\left(\Omega t, \Gamma^2\right)\right],
\end{equation}
thus the period of the coefficient matrix $\boldsymbol{A}(t)$ is half the period of the Jacobi elliptic cosine function, that is $T_{\boldsymbol{A}}=T_{\rm cn}/2$. Based on the Floquet theory \cite{Teschl2012}, it is known that $\boldsymbol{W}$ need not be periodic, however is must be of the form
\begin{equation}
 \boldsymbol{W}(t)\sim e^{\mu t}\boldsymbol{S}(t)
\end{equation}
where $\boldsymbol{S}(t)$ has period $T_{\boldsymbol{A}}$, and $\mu$ is known as a Floquet exponent (also characteristic exponent) which may be a complex. Here it has 2 such $\mu_j$ and together they satisfy
\begin{equation}\label{eq_Mu0}
  e^{\mu_1T_{\boldsymbol{A}}}\cdot e^{\mu_2T_{\boldsymbol{A}}}
  =e^{\int_{0}^{T_{\boldsymbol{A}}}{\rm tr}(\mathbf{A}(\tau))d\tau}=1,~{\rm namely,} \quad \mu_1+\mu_2=0.
\end{equation}
By choosing the initial conditions
\begin{equation}\label{eq_W0}
  \boldsymbol{W}^{(1)}(0)=\left[
\begin{array}{c}
1\\
0
\end{array}\right],~{\rm and}\quad
  \boldsymbol{W}^{(2)}(0)=\left[
\begin{array}{c}
0\\
1
\end{array}\right],
\end{equation}
we obtain solutions of the form
\begin{equation}\label{eq_Wss}
  \boldsymbol{W}^{(1)}(t)=\left[
\begin{array}{c}
W_1^{(1)}(t)\\
W_2^{(1)}(t)
\end{array}\right],~{\rm and}\quad
  \boldsymbol{W}^{(2)}(t)=\left[
\begin{array}{c}
W_1^{(2)}(t)\\
W_2^{(2)}(t)
\end{array}\right].
\end{equation}
Let $\boldsymbol{X}(t)$ is the $\left( 2\times2 \right)$ fundamental matrix of the system of Eq.~(\ref{eq_W}), as a result, we have chosen $\boldsymbol{X}(0)=\left[\boldsymbol{W}^{(1)}(0)\quad \boldsymbol{W}^{(2)}(0)\right]=\boldsymbol{I}$, then $\boldsymbol{X}(t)$ is the principal fundamental matrix. Thus we obtain the eigenvalues $\rho_1$ and $\rho_2$ of $\boldsymbol{X}(T_{\boldsymbol{A}})$, which are known as Floquet multipliers (also characteristic multipliers) satisfying
\begin{equation}\label{eq_Rho}
  \rho_{1}=e^{\mu_1T_{\boldsymbol{A}}},~{\rm and}\quad \rho_{2}=e^{\mu_2T_{\boldsymbol{A}}}.
\end{equation}
With the help of Eq.~(\ref{eq_Mu0}), we have
\begin{equation}\label{eq_Rho_Mu}
\cases{
  \rho_{1}\cdot\rho_{2}=1,&\\
  \rho_{1}+\rho_{2}={\rm tr}[\boldsymbol{X}(T_{\boldsymbol{A}})]=W_1^{(1)}(T_{\boldsymbol{A}})+W_2^{(2)}(T_{\boldsymbol{A}}).}
\end{equation}
Without losing generality, we assume $\rho_1\geq \rho_2$. Let $\phi={\rm tr}[\boldsymbol{X}(T_{\boldsymbol{A}})]/2$, we obtain
\begin{equation}\label{eq_rho_phi}
  \rho=\phi\pm\sqrt{\phi^2-1},~{\rm and}\qquad \mu=\frac{\ln(\rho)}{T_{\boldsymbol{A}}}.
\end{equation}
Different values of $\phi$ will lead to the following different situations \cite{Chicone2006}:
\begin{itemize}
  \item If $\phi>1$, then $\rho_1>1>\rho_2>0$, so that $\mu_1=-\mu_2=\frac{\ln(\rho_1)}{T_{\boldsymbol{A}}}>0$, we have an unstable solution of the form
      \begin{equation}
        \boldsymbol{W}(t)=c_1 e^{\mu_1 t}\boldsymbol{S}_1(t)+c_2 e^{-\mu_1 t}\boldsymbol{S}_2(t).
      \end{equation}
  \item If $\phi=1$, then $\rho_1=\rho_2=1$, i.e., $\mu_1=\mu_2=0$, we have an unstable solution of the form
      \begin{equation}
        \boldsymbol{W}(t)=(c_1+tc_2)\boldsymbol{S}_1(t)+c_2 \boldsymbol{S}_2(t).
      \end{equation}
  \item If $-1<\phi<1$, then we define $\eta$ by $\phi=\cos(\eta T_{\boldsymbol{A}})$ and supposing $0<\eta T_{\boldsymbol{A}}<\pi$, thus we have $\rho=e^{\pm i \eta T_{\boldsymbol{A}}}$, and $\mu=\pm i\eta T_{\boldsymbol{A}}$, we get a stable pseudo-periodic solution of the form
      \begin{equation}\label{eq_perodicS}
        \boldsymbol{W}(t)=c_1 \Re[e^{i\eta t}\boldsymbol{S}(t)]+c_2 \Im[e^{i\eta t}\boldsymbol{S}(t)].
      \end{equation}
  \item If $\phi=-1$, then $\rho_1=\rho_2=-1$, thus $\mu$ must be a pure imaginary number, i.e., $\Re(\mu)=0$, and $\mu_1=\mu_2=\frac{i\pi}{T_{\boldsymbol{A}}}$, we have an unstable solution of the form
      \begin{equation}
        \boldsymbol{W}(t)=\left[(c_1+tc_2)\boldsymbol{S}_1(t)+c_2 \boldsymbol{S}_2(t)\right]e^{\frac{i\pi t}{T_{\boldsymbol{A}}}}.
      \end{equation}
  \item If $\phi<-1$, then $\rho_2<-1<\rho_1<0$, thus $\mu$ must be a complex with algebraic form $\mu_1=-\mu_2=\zeta+\frac{i\pi}{T_{\boldsymbol{A}}}$, where $\zeta=\frac{\ln(|\rho_1|)}{T_{\boldsymbol{A}}}$, we obtain an unstable solution of the form
      \begin{equation}\label{eq_expS}
        \boldsymbol{W}(t)=\left[c_1 e^{\zeta t}\boldsymbol{S}_1(t)+c_2 e^{-\zeta t}\boldsymbol{S}_2(t)\right]e^{\frac{i\pi t}{T_{\boldsymbol{A}}}}.
      \end{equation}
\end{itemize}
In above cases, the $\boldsymbol{S}(t)$, $\boldsymbol{S}_1(t)$, and $\boldsymbol{S}_2(t)$ are all periodic with period $T_{\boldsymbol{A}}$.

In short, the instability threshold $\lambda_c$ of the system can be obtained by solving $|\phi|=1$. Besides, as long as the real part of one of the Floquet exponents becomes positive; i.e., $\Re(\mu)>0$, the amplitude of the error will increase exponentially, so that the system is unstable, and the characteristic time scale of instability of system could be estimated as
\begin{equation}\label{eq_T_beta}
  T\propto \frac{1}{|\Re(\mu)|}.
\end{equation}
In principle, if we integrate differential equation (\ref{eq_W}) and get the solution (\ref{eq_Wss}), we can get the complete (exact) solution to the problem. Unfortunately, it is difficult for us to get an analytical solution (i.e., explicit expression) due to the complexity of related calculations of the Jacobian elliptic cosine function, in main text we mainly apply numerical integration to solve the differential equation (\ref{eq_W}). Nevertheless, to facilitate the description of the subsequent related results, we call the results obtained by numerical integration as the \emph{exact} ones.

In addition, we expand the Jacobian elliptic cosine function by trigonometric series and get an \emph{approximate} solution of $\lambda_c$, namely
\begin{eqnarray}\label{eq_Lc_Cut_beta}
\lambda_c = &\frac{\pi }{3 N}+\frac{(6+11\pi)\pi}{18 N^2}+\frac{(576+1056\pi+2543 \pi ^2)\pi}{1728 N^3}\nonumber\\
&+\frac{(1728+3168 \pi+7629 \pi ^2+23222 \pi ^3-78 \pi ^5)\pi}{5184 N^4}+O\left(\frac{1}{N^5}\right),
\end{eqnarray}
which means that
\begin{equation}\label{eq_LcVsN_Beta}
  \lambda_c\propto N^{-1}
\end{equation}
in the large $N$ limit. Besides, as an illustrating example, for the fixed system size $N=32$, we obtain
\begin{equation}\label{eq_ReU2_beta_3order}
  |\Re(\mu)|= 0.2263(\lambda-\lambda_c)^{1/2}-0.4760(\lambda-\lambda_c)^{3/2}+\cdots,
\end{equation}
which means that near the instability threshold, the stability time follows
\begin{equation}\label{eq_T_beta2}
  T\propto (\lambda-\lambda_c)^{-1/2},
\end{equation}
which has been checked that it is size independent. Expressions (\ref{eq_Lc_Cut_beta}) to (\ref{eq_T_beta2}) are our main theoretical results for the FPUT-$\beta$ model, see Appendix B for a detailed derivation of these expressions.

\begin{figure}[t]
  \centering
  \includegraphics[width=1\textwidth]{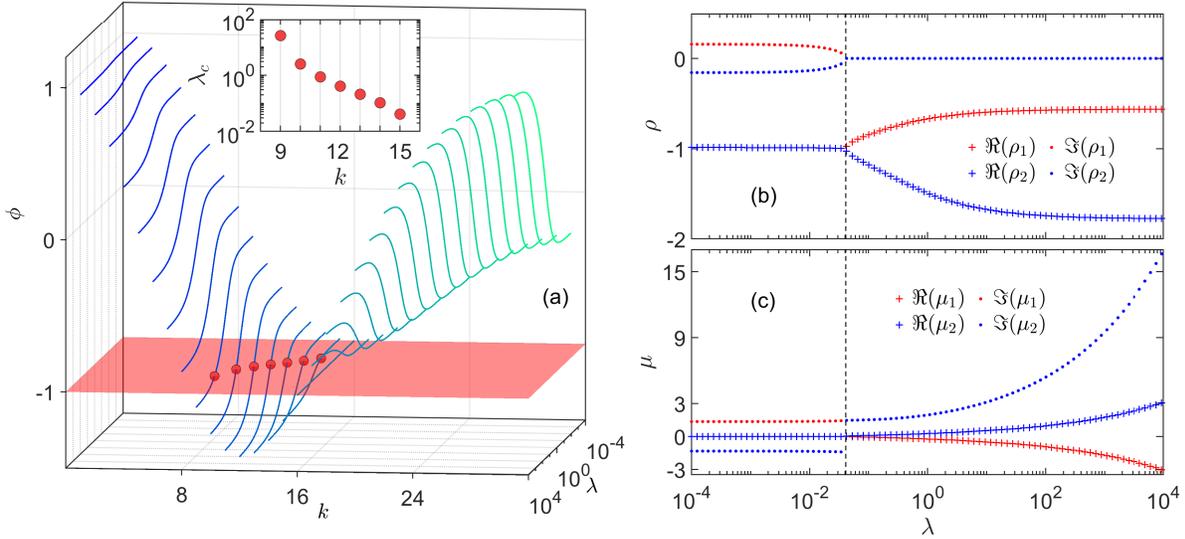}
  \caption{(a) The dependence of $\phi$ on the wave number $k$ (where $1\leq k\leq N-1$) and the perturbation strength $\lambda$ for the fixed size of system $N=32$. The intersection of curve and the plane $\phi=-1$ gives the instability threshold $\lambda_c$ of a mode, which is plotted in the inset. (b) The Floquet multiplier $\rho$ as a function of $\lambda$ for $k=15$ in panel (a). The crosses and the dots denote the real and imaginary parts, respectively. (c) The Floquet exponent $\mu$ as a function of $\lambda$, which is directly obtained from the panel (b) by applying Eq.~(\ref{eq_Rho}). The vertical dashed lines in panels (b) and (c) are for $\lambda_c$.}\label{figPhi}
\end{figure}

Figure~\ref{figPhi}(a) shows the exact solution of $\phi$ as a function of the wave number $k$ and the perturbation strength $\lambda$ for the fixed system size $N=32$. It is clearly seen that $\phi<1$ for all $k$ in our calculations $\lambda\in[10^{-4},10^4]$. The function curve of $\phi$ of only a few modes crosses the plane $\phi=-1$, which gives the instability threshold $\lambda_c$ of the corresponding mode (see inset in the main panel). From the inset in Fig.~\ref{figPhi}(a), it can be seen that the $\lambda_c$ of the $(N/2-1)$ mode is the smallest, which means that it will lose stability first. This result is completely consistent with the phenomenon observed in our MDS (see Fig.~\ref{figEkT}(a) in Sec.~\ref{sec:4}). At the same time, it means that the instability threshold of the whole system is determined by instability of the $(N/2-1)$ mode. Therefore, we next mainly focus on the behavior of the Floquet multiplier $\rho$ and the Floquet exponent $\mu$ of the $(N/2-1)$ mode.

Figures~\ref{figPhi}(b) and \ref{figPhi}(c) show, respectively, the Floquet multiplier $\rho$ and the Floquet exponent $\mu$ as functions of the perturbation strength $\lambda$ for the $(N/2-1)$ mode. Note that $\rho_{j}$ are complex numbers, and $\mu_{j}$ are pure imaginary numbers (i.e., $\Re(\mu_1)=\Re(\mu_2)=0$) when $\lambda<\lambda_c$, thus the evolution of the error of the mode takes the form of Eq.~(\ref{eq_perodicS}), i.e., periodic form. On the contrary, $\rho_{j}$ become real numbers and $\rho_2<-1<\rho_1<0$, then $\mu_{j}$ become complex numbers (i.e., $\Re(\mu_2)=-\Re(\mu_1)>0$) when $\lambda>\lambda_c$, so that the amplitude of the error will increase exponentially with the form of Eq.~(\ref{eq_expS}), that is, the system loses stability. The characteristic time scale is given by $\Re(\mu_2)$, and estimated as expression (\ref{eq_T_beta2}) for small $\lambda$ (i.e., $0<\lambda-\lambda_c\ll1$).

\begin{figure}[t]
  \centering
  \includegraphics[width=1\textwidth]{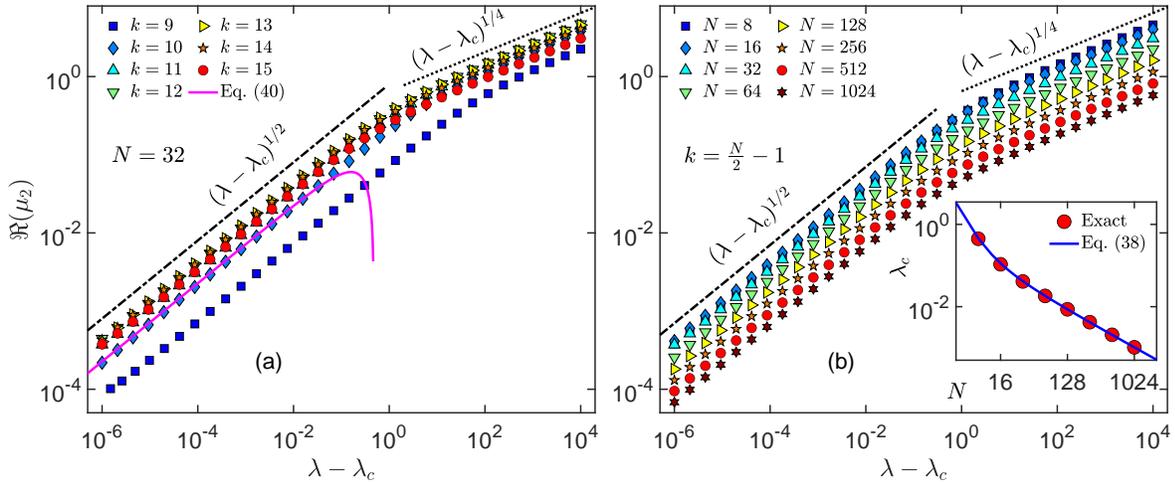}
  \caption{(a) $\Re(\mu_2)$ as a function of $(\lambda-\lambda_c)$ for various wave number $k$ at the fixed $N=32$ in log-log scale. The magenta solid is the function curve of expression (\ref{eq_ReU2_beta_3order}). The lines of dashed-dotted and dotted, with different slope, are drawn for reference. (b) The same as in panel (a) but is for different system size $N$, and $k={N}/{2}-1$. Inset: The instability threshold $\lambda_c$ as a function of $N$ in log-log. The blue solid line is the function curve of expression (\ref{eq_Lc_Cut_beta}).}\label{figMu}
\end{figure}

In Fig.~\ref{figMu}(a), we present $\Re(\mu_2)$ as a function of $(\lambda-\lambda_c)$ at the fixed $N=32$ for various wave numbers. Note that for all modes presented, the two scaling relationships can be clearly recognized, that is, $\Re(\mu_2)\propto (\lambda-\lambda_c)^{1/2}$ is for the smaller $\lambda$, while $\Re(\mu_2)\propto (\lambda-\lambda_c)^{1/4}$ is for the larger $\lambda$. Notice that for $(\lambda-\lambda_c)<1$, the result here, i.e., Eq.~(\ref{eq_ReU2_beta_3order}), agrees qualitatively with the exact result obtained by numerical integration. From Fig.~\ref{figMu}(a), it also can be seen that there is a large deviation in the results for $\lambda>1$. We conjecture that this is caused by the finite truncation of the trigonometric series expansion of the Jacobian elliptic cosine function. From Eq.~(\ref{13}), we know that $\Gamma^2$ increases monotonically with the increase of $\lambda$, so that one needs to keep more terms to achieve the desired computational accuracy especially when $\lambda>1$. Of course, this truncation treatment may also be the source of quantitative differences in the results when $\lambda<1$.

​To test whether these scaling relationships are the specific behavior for $N=32$, we give the results for various $N$ at the fixed $k=(N/2-1)$ in Fig.~\ref{figMu}(b). It can be seen that the scaling behavior is size independent. Besides, in Fig.~\ref{figMu}(b), we show the dependence of the instability threshold $\lambda_c$ on the system size $N$ (see red dots in the inset), where the approximate solution (blue solid line) and the exact solution (red dots) agree very well.

\subsection{Analysis for the FPUT-$\alpha$ chains}

For the FPUT-$\alpha$ chain, Eq.~(\ref{17}) can be formulated as
\begin{equation}\label{20}
\Delta \ddot{Q}_{k}=-\omega_{k}^{2} \Delta Q_{k}-\frac{2 \alpha }{\sqrt{2 N}} \omega_{k}\omega_{\frac{N}{2}} Q_{\frac{N}{2}} \sum_{k_{3}} \omega_{k_{3}} C_{k, \frac{N}{2}, k_{3}} \Delta Q_{k_{3}},
\end{equation}
where the coefficient
\begin{equation}\label{21}
C_{k, \frac{N}{2}, k_{3}}=\delta_{k - \frac{N}{2} - k_{3}, 0}+\delta_{k - \frac{N}{2} + k_{3}, 0}+\delta_{k + \frac{N}{2} - k_{3}, 0}-\delta_{k + \frac{N}{2} + k_{3}, 2N}.
\end{equation}
From Eq.~(\ref{21}), we note that the coefficient does not vanish only when $k_3=N/2-k$ or $k_3=N/2+k$ for $0<k<N/2$ while $k_3=k-N/2$ or $k_3=3N/2-k$ for $N/2<k<N$. After a careful analysis (details see Appendix C), the evolution of the error controlled by Eq.~(\ref{20}) only has the following three types:

$\bullet$ \emph{The one-mode equation}: If $k=N/2$, from Eq.~(\ref{20}) we have
\begin{equation}\label{22}
\Delta \ddot{Q}_{\frac{N}{2}}=-\omega_{\frac{N}{2}}^{2} \Delta Q_{\frac{N}{2}}.
\end{equation}
The solution of Eq.~(\ref{22}) can be easily obtained:
\begin{equation}\label{23}
\Delta Q_{\frac{N}{2}}(t)=A \cos \left(\sqrt{2} t\right),
\end{equation}
where $\omega_{\frac{N}{2}}=\sqrt{2}$ has be substituted, and $A$ is a free parameter determined by the initial condition of the error. Namely, the error of the $N/2$ mode is decoupled with other modes and evolves in the form of oscillation, i.e., the one-mode equation of error is stable.

$\bullet$ \emph{The two-mode coupled equations}: If $k=N/4$ or $k=3N/4$, Eq.~(\ref{20}) can be specifically written as
\begin{equation}\label{24}
\cases{\Delta \ddot{Q}_{\frac{N}{4}}=-\omega_{\frac{N}{4}}^{2}
\left(1+\frac{2\alpha }{\sqrt{N}}Q_{\frac{N}{2}}\right)
 \Delta Q_{\frac{N}{4}}-
 \frac{2\alpha }{\sqrt{N}} \omega_{\frac{N}{4}} \omega_{\frac{3N}{4}} Q_{\frac{N}{2}} \Delta Q_{\frac{3N}{4}},&~\\
\Delta \ddot{Q}_{\frac{3N}{4}}=-\omega_{\frac{3N}{4}}^{2}
\left(1-\frac{2\alpha }{\sqrt{N}}Q_{\frac{N}{2}}\right)
 \Delta Q_{\frac{3N}{4}}-
 \frac{2\alpha }{\sqrt{N}} \omega_{\frac{N}{4}} \omega_{\frac{3N}{4}} Q_{\frac{N}{2}} \Delta Q_{\frac{N}{4}}.
&~\\}
\end{equation}


$\bullet$ \emph{The four-mode coupled equations}: The above three modes $k=N/4$, $N/2$, and $3N/4$ divide the space of $k$ into four intervals: $(0,N/4)$, $(N/4,N/2)$, $(N/2,3N/4)$, and $(3N/4,N)$. We have verified that in the four cases one obtains the same results (see analysis in Appendix C). Let us consider the case of $k\in(0,N/4)$. From Eq.~(\ref{20}), we obtain
\begin{equation}\label{26}
\cases{\Delta \ddot{Q}_{k}=-\omega_{k}^{2} \Delta Q_{k}-\frac{2 \alpha\omega_{k}}{\sqrt{N}}  Q_{\frac{N}{2}}\mathcal{C}_1,&~\\
\Delta \ddot{Q}_{\frac{N}{2}-k}=-\omega_{\frac{N}{2}-k}^{2} \Delta Q_{\frac{N}{2}-k}-\frac{2 \alpha\omega_{\frac{N}{2}-k} }{\sqrt{N}} Q_{\frac{N}{2}}\mathcal{C}_2,&~\\
\Delta \ddot{Q}_{\frac{N}{2}+k}=-\omega_{\frac{N}{2}+k}^{2} \Delta Q_{\frac{N}{2}+k}-\frac{2 \alpha\omega_{\frac{N}{2}+k}}{\sqrt{N}}  Q_{\frac{N}{2}}\mathcal{C}_3,&\\
\Delta \ddot{Q}_{N-k}=-\omega_{N-k}^{2} \Delta Q_{N-k}-\frac{2 \alpha\omega_{N-k}}{\sqrt{N}}  Q_{\frac{N}{2}}\mathcal{C}_4,&
}
\end{equation}
where $\mathcal{C}_1=\omega_{\frac{N}{2}-k} \Delta Q_{\frac{N}{2}-k}+\omega_{\frac{N}{2}+k} \Delta Q_{\frac{N}{2}+k}$, $\mathcal{C}_2=\omega_{k} \Delta Q_{k}+\omega_{N-k} \Delta Q_{N-k}$, $\mathcal{C}_3=\omega_{k} \Delta Q_{k}-\omega_{N-k} \Delta Q_{N-k}$, and $\mathcal{C}_4=\omega_{\frac{N}{2}-k} \Delta Q_{\frac{N}{2}-k}-\omega_{\frac{N}{2}+k} \Delta Q_{\frac{N}{2}+k}$. Note that the four-mode coupled equations (\ref{26}) completely degenerate into the two-mode coupled equations (\ref{24}) when $k=N/4$ or $k=3N/4$. Hence, we take the four-mode coupled equations as an example to make a brief illustration in the following.

In a similar way, the above coupled equations (\ref{26}) can be further rewritten in the form of matrix, and the coefficient matrix is a ($8\times8$) periodic one. The period $T$ of the coefficient matrix is determined by the $N/2$ mode (see Eq.~(\ref{10})), that is, $T=\sqrt{2}\pi$. Let $\boldsymbol{X}(t)$ is the $\left( 8\times8 \right)$ fundamental matrix of Eq.~(\ref{26}), and if we choose $\boldsymbol{X}(0)=\boldsymbol{I}$, then $\boldsymbol{X}(t)$ is the principal fundamental matrix. Thus the Floquet multipliers, i.e., eigenvalues $\rho_j$ ($j=1, 2, \cdots, 8$) of $\boldsymbol{X}(T)$ can be obtained directly through diagonalization, and the Floquet exponents $\mu_j={\ln(\rho_j)}/{T}$. Similarly, here we use numerical integration to solve the differential equations to find the exact solutions. As a comparison, the approximate solution of $\lambda_c$ (details see Appendix D) is shown as
\begin{equation}\label{eq_Lc_alpha}
\lambda_c=\frac{0.33}{N}-\frac{0.86}{N^{2}}+\frac{0.48}{N^{3}}+O\left(\frac{1}{N^4}\right),
\end{equation}
which also gives that $\lambda_c\propto N^{-1}$ in the large $N$ limit. Unfortunately, we did not obtain the approximate analytical solution of $\mu_j$ since the calculation related to the FPUT-$\alpha$ model is very complicated, we only give the numerical (exact) result of $\mu_j$ below.

\begin{figure}[t]
  \centering
  \includegraphics[width=1\textwidth]{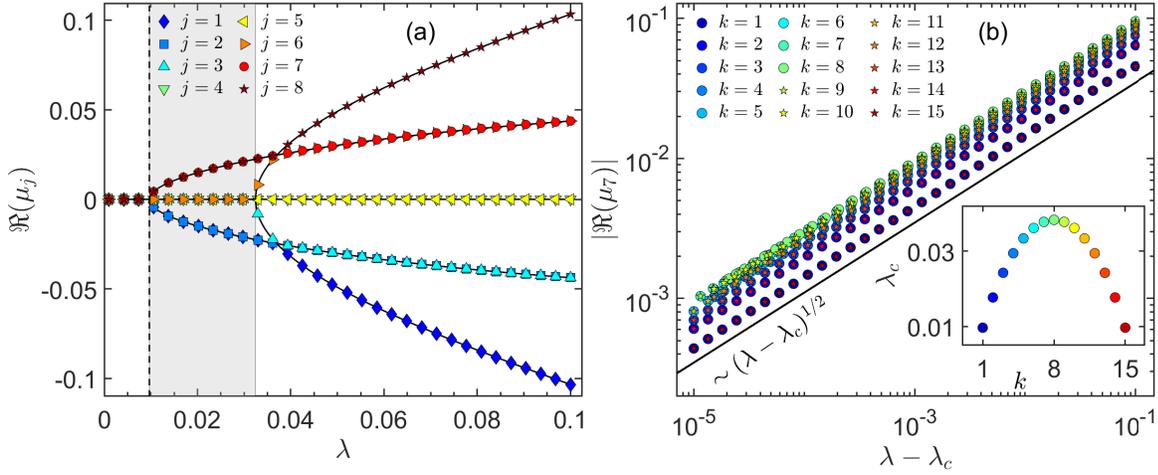}\\
  \caption{(a) $\Re(\mu_j)$ as a function of perturbation strength $\lambda$ for the four-mode coupled equations~(\ref{26}) where $k=1$ is adopted at the fixed size $N=32$. The vertical dashed lines is for $\lambda_c$. (b) The dependence of $|\Re(\mu_7)|$ on $(\lambda-\lambda_c)$ for $k\in(0,N/2)$ at the fixed size $N=32$, in log-log scale. The black solid line is drawn for reference. Inset: the dependence of the instability threshold $\lambda_c$ on the wave number $k$.}\label{figAlphaN32}
\end{figure}

Figure~\ref{figAlphaN32}(a) shows $\Re(\mu_j)$ as a function of $\lambda$ at the fixed system size $N=32$ for setting $k=1$ in Eq.~(\ref{26}). In our numerical calculations, the output of $\mu_j$ is sorted from smallest to largest in real part. Note that in the whole range of $\lambda$ calculated, $\Re(\mu_8)=-\Re(\mu_1)\geq0$, $\Re(\mu_7)=-\Re(\mu_2)\geq0$, $\Re(\mu_6)=-\Re(\mu_3)\geq0$, and $\Re(\mu_5)=-\Re(\mu_4)\geq0$. The transition point of $\Re(\mu)$ from zero to non-zero gives the instability threshold $\lambda_c$ (see vertical dashed line). It can be seen that when $\lambda<\lambda_c$, $\Re(\mu_j)=0$ for $j\in[1,8]$. On the contrary, when $\lambda>\lambda_c$, $\Re(\mu_7)=-\Re(\mu_2)=\Re(\mu_8)=-\Re(\mu_1)>0$ while $\Re(\mu_3)=\Re(\mu_4)=\Re(\mu_5)=\Re(\mu_6)=0$ (see $\lambda$ within gray area). As $\lambda$ increases further, $\Re(\mu_8)=-\Re(\mu_1)>\Re(\mu_7)=-\Re(\mu_2)=\Re(\mu_6)=-\Re(\mu_3)>0$ while $\Re(\mu_4)=\Re(\mu_5)=0$. Through comprehensive comparison, we notice that $\Re(\mu_7)$  changes from zero to non-zero (i.e., which gives the $\lambda_c$) first, and is always on the same curve. Hence, we focus on $\Re(\mu_7)$ in the following.

In Fig.~\ref{figAlphaN32}(b), we show the dependence of $|\Re(\mu_7)|$ on $(\lambda-\lambda_c)$ for various $k$ at the fixed $N=32$. It is clearly seen that $\Re(\mu_7)\propto (\lambda-\lambda_c)^{1/2}$, which is the same as the result of the FPUT-$\beta$ model for small $\lambda$. We also show the instability threshold $\lambda_c$ as a function of wave number $k$ in the inset of Fig.~\ref{figAlphaN32}(b). It can be seen that $\lambda_c$ is nonmonotonic and completely symmetric with respect to $k={N}/{4}=8$. Similarly, the threshold corresponding to the $(N/2-1)$ mode is the smallest, that is, it will lose stability first. This is completely consistent with the phenomenon observed in MDS for the FPUT-$\alpha$ model (see Fig.~\ref{figEkT}(b)).

\section{\label{sec:4}Results of molecular dynamics simulation}

\subsection{The results for the FPUT-$\beta$ chains}

\begin{figure}[t]
  \centering
  \includegraphics[width=1\columnwidth]{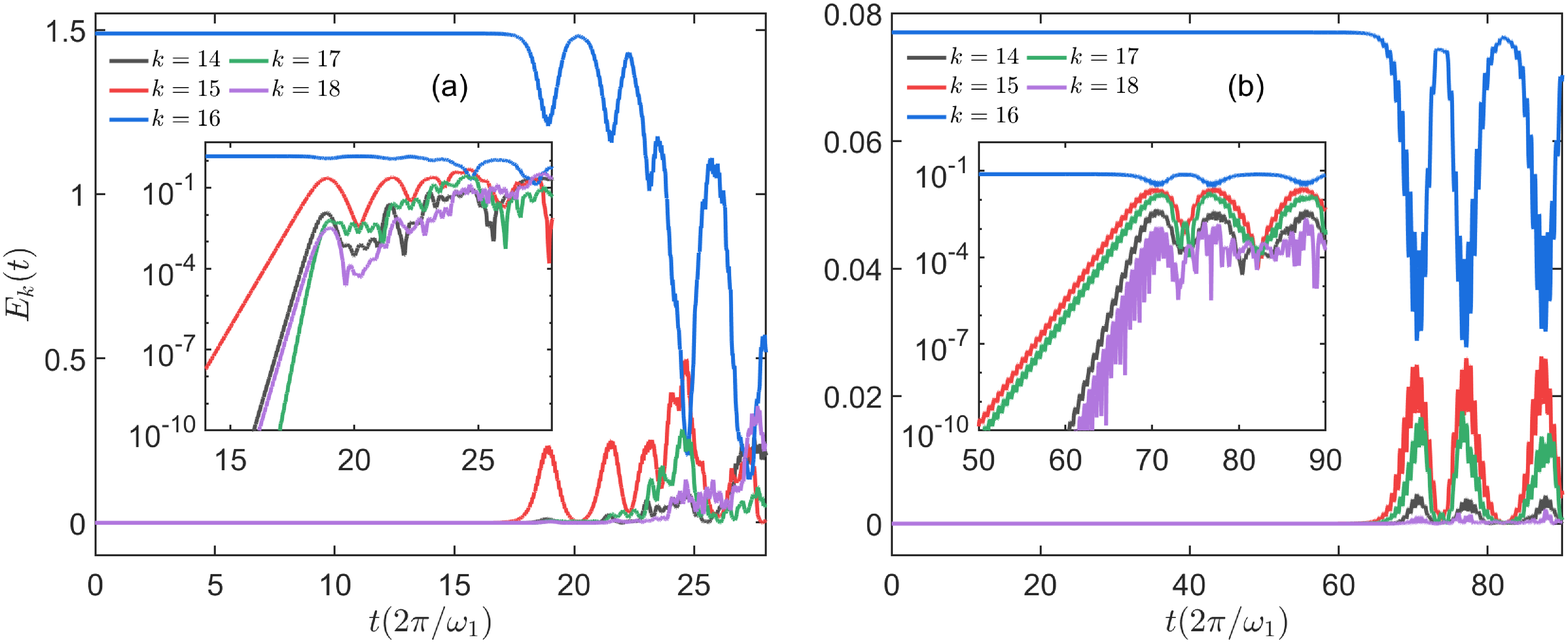}\\
  \caption{(a) The energy $E_k(t)$ versus time $t$ for the FPUT-$\beta$ model while the $N/2$ mode was excited initially. The system size $N=32$, the nonlinear coupling strength $\beta=1$, and the total energy $E=1.488$ thus energy density $\varepsilon=E/(N-1)$ are fixed, i.e., $\lambda=\beta\varepsilon=0.048$. (b) Same as panel (a) but for the FPUT-$\alpha$ model. The system size $N=32$, the nonlinear coupling strength $\alpha=2.25$, and the total energy $E=0.077$ thus perturbation strength $\lambda=\alpha^2\varepsilon\approx0.0126$. Insets: Same as the main panels but in semi-log scale.}\label{figEkT}
\end{figure}

The $N/2$ mode acting as a one-mode solution of the system will be unstable when the intrinsic nonlinear perturbation of the system reaches a certain degree. To study the instability dynamics of this mode numerically, the equations of motion (\ref{eqMotion}) are integrated by the eighth-order Yoshida method \cite{YOSHIDA1990262}. The typical time step $\Delta t=0.01$; the corresponding relative error in energy conservation is far less than $10^{-6}$.

In Fig~\ref{figEkT}(a) and \ref{figEkT}(b), we track the evolution of the energy of the $N/2$ mode and its several adjacent modes with time under the $N/2$ mode excited initially, i.e, $E_{k}(0)=(N-1)\varepsilon\delta_{\frac{N}{2},k}$, for the FPUT-$\beta$ and the FPUT-$\alpha$ chains, respectively. In other words, we did not artificially introduce errors on the other modes in our MSD. The error mentioned in the section of theoretical analysis is inevitably introduced by the truncation error of computer and the error of the integration algorithm in simulations. It is clearly seen that the $N/2$ mode becomes unstable after a long period of time for the two models. At the same time, the (${N}/{2}-1$) mode gains energy first and grows exponentially (see insets in main panels). It should be pointed out that the instability of the single-mode solution is an intrinsic property ruled by the system parameters but not caused by the accuracy of the numerical algorithm.

To study the relationship between the stability of the $N/2$ mode and the perturbation strength quantitatively, we define the stability time $T$ at the moment when the energy loss of the $N/2$ mode reaches a certain threshold for the first time; i.e., $\Delta E_{\frac{N}{2}}=[E_{\frac{N}{2}}(0)-E_{\frac{N}{2}}(T)]/E_{\frac{N}{2}}(0)=\xi$, where $\xi$ is a free parameter that controls the threshold.

\begin{figure}[t]
\centering
\includegraphics[width=1\textwidth]{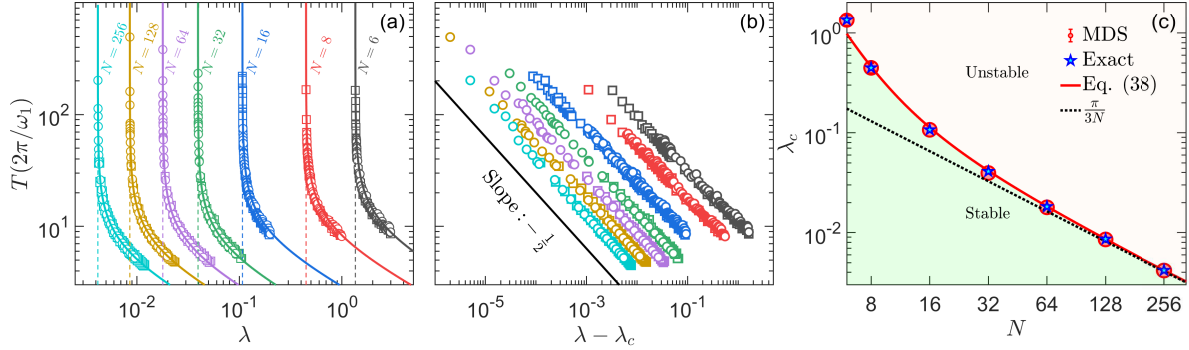}
\caption{(a) The stability time $T$ versus perturbation strength $\lambda$ for the FPUT-$\beta$ model with different system size $N$ in log-log scale. The circles and squares are the numerical results for the threshold value $\xi=0.01$ and $\xi=0.001$, respectively. The solid lines are for $T\sim(\lambda-\lambda_c)^{-1/2}$, which are plotted for reference, and the vertical dashed lines correspond to $\lambda_c$. Panel (b): the same as in (a) but shown as a function of $(\lambda-\lambda_c)$. The solid lines with slope $-\frac{1}{2}$ are drawn for reference. (c) The dependence of $\lambda_c$ on $N$ in log-log scale for the $N/2$ mode of the FPUT-$\beta$ model. The blue dotted line is an asymptote for large $N$, i.e., the first term of Eq.~(\ref{eq_Lc_Cut_beta}).}\label{figTBeta}
\end{figure}

Figure~\ref{figTBeta} (a) shows the numerical results (plotted by the scatter points) of $T$ as a function of $\lambda$ for a given initial condition $\varphi_{\frac{N}{2}}=0$. The solid lines are fitting curves based on expression (\ref{eq_T_beta2}); and the vertical dashed lines are corresponding to $\lambda_c$ which gives the instability threshold of the $N/2$ mode for a certain $N$. It can be seen that the data points are well covered by the fitting curves. To clearly show the dependence of $T$ on the $\lambda$, we present the relationship between $T$ and $(\lambda-\lambda_c)$ in log-log scale in Fig.~\ref{figTBeta} (b). Note that all the data points fall onto the lines with slope $-\frac{1}{2}$, suggesting that again, $T\propto(\lambda-\lambda_c)^{-\frac{1}{2}}$ is confirmed convincingly. In our MDS, the variation of the perturbation strength is realized by changing the nonlinear coupling coefficient with fixed energy density $\varepsilon=0.01$, and this strategy is used throughout for all the numerical results presented. Besides, $\xi=0.01$ is adopted for controlling the threshold value. Though assuming the threshold value is artificial, it does not influence the whole behavior of $T$ vs $\lambda$, especially the estimation of $\lambda_c$, which has been checked by $\xi=0.001$ (see squares in Figs.~\ref{figTBeta} (a) and \ref{figTBeta} (b)).

Next, we will explore the dependence of $\lambda_c$ on $N$. In fact, the initial excitation phase of the $N/2$ mode has a significant influence on the instability threshold. To eliminate this effect, the instability thresholds for different system sizes are the averages which were done over 24 phases uniformly distributed in $\left[ 0,2\pi \right]$. Figure~\ref{figTBeta} (c) presents the numerical results of $\lambda_c$ as a function of $N$. It can be seen that the numerical results and theoretical ones agree well.

\subsection{The results for the FPUT-$\alpha$ chains}

\begin{figure}[t]
	\centering
    \includegraphics[width=1\textwidth]{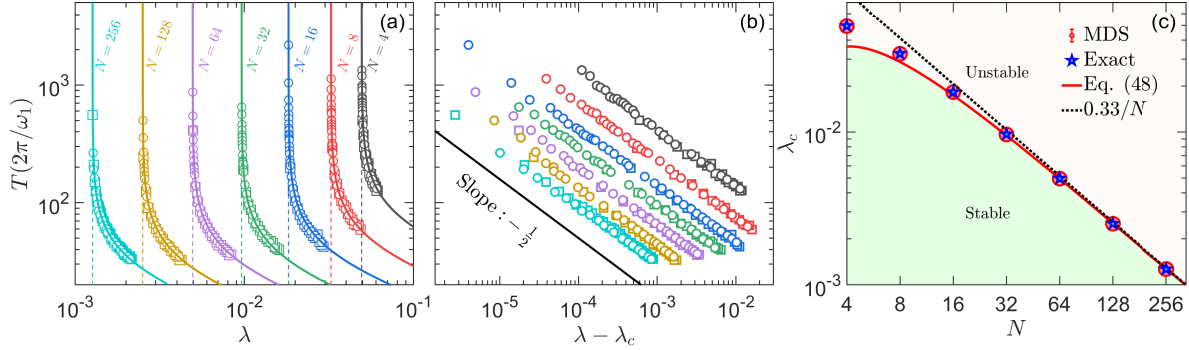}
	\caption{(a) The stability time $T$ as a function $\lambda$ for the FPUT-$\alpha$ model with different system size $N$ in log-log scale. The circles and squares are the numerical results for the threshold value $\xi=0.01$ and $\xi=0.001$, respectively. The solid lines are for $T\sim(\lambda-\lambda_c)^{-1/2}$, which are plotted for reference, and the vertical dashed lines correspond to $\lambda_c$. Panel (b): the same as in (a) but shown as a function of $(\lambda-\lambda_c)$. The solid lines with slope $-\frac{1}{2}$ are drawn for reference. (c) The dependence of $\lambda_c$ on $N$ in log-log scale for the $N/2$ mode of the FPUT-$\alpha$ model. The blue dotted line is an asymptote for large $N$, i.e., the first term of Eq.~(\ref{eq_Lc_alpha}).}
\label{figTalpha}
\end{figure}

The numerical method in the research of the FPUT-$\beta$ model is fully applied to the study of the FPUT-$\alpha$ model. Figure~\ref{figTalpha} (a) shows the stability time $T$ as a function of $\lambda$ for the FPUT-$\alpha$ model. The solid lines are fitting curves based on the expression (\ref{eq_T_beta2}); and the vertical dashed lines also give the instability threshold $\lambda_c$. Figure~\ref{figTalpha} (b) presents the dependence of $T$ on $(\lambda-\lambda_c)$ in log-log scale. It is clearly seen that $T\propto(\lambda-\lambda_c)^{-\frac{1}{2}}$.

Figure~\ref{figTalpha} (c) displays the results of $\lambda_c$ as a function of $N$ for the FPUT-$\alpha$ model. It can be seen that the numerical simulation results agree with the exact results very well, but the approximate analytical solution (see the red solid line, i.e, Eq.~(\ref{eq_Lc_alpha}), detailed analysis see Appendix D) is estimated to be smaller in small size, while the three are completely consistent for larger $N$.

Comparing the results of the FPUT-$\alpha$ model (see again Fig.~\ref{figTalpha} (c)) and the FPUT-$\beta$ model (see again Fig.~\ref{figTBeta} (c)), it is seen that except for the qualitative difference of the law of $\lambda_c$ in the range of small size, the behavior in the large size is qualitatively identical; i.e., $\lambda_c\propto N^{-1}$. It should be noticed that the emergence of this unified law is that the perturbation strength in the FPUT-$\alpha$ model is defined as $\lambda=\alpha^2\varepsilon$ rather than $\lambda=\alpha\sqrt{\varepsilon}$.

\section{\label{sec:5} Conclusions and discussions}

In summary, we have studied the instability dynamics of the $N/2$ mode theoretically and numerically in the FPUT-$\beta$ and -$\alpha$ models under fixed boundary conditions. The dependence of $\lambda_c$ on $N$ is analytically obtained for both models; i.e., Eq.~(\ref{eq_Lc_Cut_beta}) for the former, and Eq.~(\ref{eq_Lc_alpha}) for the latter, which agree well with the exact solution of numerical integration and the results of MDS. It is found that $\lambda_c\propto N^{-1}$ in the large $N$ limit for both models; in addition, in both models, the stability time, $T$, as a function of $\lambda$ follows same behavior $T\propto (\lambda-\lambda_c)^{-\frac{1}{2}}$, which is size independent. Our results fully show that the $N/2$ mode in the two models has exactly the same instability dynamic behavior.


It is worth noting that the total number of NNMs is very limited, thus in the large $N$ limit the probability that the initial excitation modes just select NNMs is very small (approaching zero). Besides, the instability threshold also tends to zero in the large $N$ limit. Consequently, it is expected that the thermalization dynamics of a generic Hamiltonian system in the large $N$ limit is mainly dominated by the mechanism of multi-wave interactions, that is, the universal thermalization law would be observed \cite{fu2019universal,Fu2019PRER,Pistone2018,PhysRevE.100.052102,PhysRevLett.124.186401,2020arXiv200503478W,PhysRevE.104.L032104,Feng_2022}. However, in the case of finite size, especially in the case of small $N$, the probability itself of selecting the initial excitation mode to NNMs increases, and the instability threshold of NNMs also increases. Therefore, in such case, if the excitation configuration of the system at the time of initial excitation happens to be NNMs, there will be a threshold for thermalization of the system, i.e., the system will not enter the thermalized state when the perturbation strength is lower than the threshold; on the contrary, when the perturbation strength is greater than the instability threshold, the thermalization dynamics of the system will be jointly determined by instability dynamics and multi-wave resonance mechanism. It is expected that the relationship between the thermalization time and the perturbation strength is not a pure power-law function.

\section*{Acknowledgments}
We acknowledge support by the NSFC (Grants No. 11975190, No. 12005156, No. 11975189, No. 12047501, and No. 11764035), and by the Natural Science Foundation of Gansu Province (Grants No. 20JR5RA494, and No. 21JR1RE289), and by the Innovation Fund for Colleges and Universities from Department of Education of Gansu Province (Grant No. 2020B-169), and by the Project of Fu-Xi Scientific Research Innovation Team, Tianshui Normal University (Grant No. FXD2020-02), and by the Education Project of Open Competition for the Best Candidates from Department of Education of Gansu Province, China (Grant No. 2021jyjbgs-06). We sincerely appreciate the anonymous reviewers for constructive suggestions on the final presentation of this work.

\section*{APPENDIX A: THE MOTION EQUATIONS OF THE NORMAL MODES}

\setcounter{equation}{0}
\setcounter{figure}{0}
\makeatletter
\renewcommand{\theequation}{A\arabic{equation}}
\renewcommand{\thefigure}{A\arabic{figure}}

Here we supplement some of the details needed to derive the motion equations of modes. Using Euler formula, equations (\ref{3b}) and (\ref{5}) in the main text can be rewritten as follows
\begin{equation}
x_{j}=\sqrt{\frac{2}{N}} \sum_{k=1}^{N-1}Q_{k}\sin \left(\frac{jk\pi}{N}\right)=i\sqrt{\frac{1}{2N}} \sum_{k=1}^{N-1}Q_{k}\left(e^{-\frac{ijk\pi}{N}}-e^{\frac{ijk\pi}{N}}\right),
\end{equation}
and
\begin{equation}
\omega_{k}=2 \sin \left(\frac{\pi k}{2 N}\right)=i\left(e^{\frac{-ik\pi}{2 N}}-e^{\frac{ik\pi}{2 N}}\right),\quad 1\le k\le N-1.
\end{equation}
To facilitate the following discussion on the properties of normal modes, the explicit expression is given here
\begin{equation}
Q_{k}=\sqrt{\frac{2}{N}} \sum_{j=1}^{N}x_j\sin \left(\frac{jk\pi}{N}\right).
\end{equation}
Note that
\begin{equation}
  \cases{
  Q_{2N\pm k} =\pm Q_k;\\
  \omega_{2N\pm k}  = \mp\omega_k.
  }
\end{equation}
The part of the nonlinear interaction potential in the Hamiltonian (\ref{eqHam}) can be written as
\begin{eqnarray}\label{eq_AVn}
V_n&=\sum_{j=1}^N\left(x_{j}-x_{j-1}\right)^n \nonumber \\  \nonumber
=&\left(i\sqrt{\frac{1}{2N}} \right)^n\sum_{j=1}^N\left[ \sum_{k=1}^{N-1}Q_{k}\left(e^{-\frac{ik\pi}{2N}}-e^{\frac{ik\pi}{2N}}\right)\left(e^{-\frac{i(j-1/2)k\pi}{N}}+
e^{\frac{i(j-1/2)k\pi}{N}}\right)\right]^n\\ \nonumber
=&\left(\sqrt{\frac{1}{2N}} \right)^n\sum_{j=1}^N\left[ \sum_{k=1}^{N-1}Q_{k}\omega_k\left(e^{-\frac{i(j-1/2)k\pi}{N}}+
e^{\frac{i(j-1/2)k\pi}{N}}\right)\right]^n\\
=&\left(\sqrt{\frac{1}{2N}} \right)^n \sum_{k_1,k_2,\dots,k_n}^{N-1}Q_{k_1}Q_{k_2}\dots Q_{k_n}\omega_{k_1}\omega_{k_2}\dots \omega_{k_n}D_{k_1, k_2, \dots, k_n},
\end{eqnarray}
where
\begin{eqnarray}
  D_{k_1, k_2, \dots, k_n} \nonumber
=\sum_{j=1}^{N}\Bigg[&e^{-\frac{i(j-1/2)(k_1+k_2+\dots+k_n)\pi}{N}}
+e^{-\frac{i(j-1/2)(k_1+k_2+\dots-k_n)\pi}{N}}+\\
&+\dots \nonumber \\
&+e^{-\frac{i(j-1/2)(k_1-k_2-\dots-k_n)\pi}{N}}
+e^{-\frac{i(j-1/2)(-k_1-k_2-\dots-k_n)\pi}{N}}\Bigg].
\end{eqnarray}
There are $2^n$ terms in the square brackets above. For simplicity, we set $\chi =\pm k_1\pm k_2\dots\pm k_n$, and then $\chi\in[-n(N-1),n(N-1)]$ is an integer. Applying the summation formula of geometric series, we have
\begin{equation}
S=\sum_{j=1}^Ne^{\frac{i(j-1/2)\chi\pi}{N}}=
\cases{
N,& $\chi=0$;\\
e^{\frac{i\chi\pi}{2N}}\cdot
\frac{1-e^{i\chi\pi}}{1-e^{i\chi\pi/N}}
=\frac{\sin\left(\frac{\chi\pi}{2}\right)
e^{\frac{i\chi\pi}{2}}}{\sin\left(\frac{\chi\pi}{2N}\right)},&$\chi\neq0$.
}
\end{equation}
Based on the formula of de Moivre, for $\chi\neq0$, we have
\begin{eqnarray}
 S&= \frac{\sin\left(\frac{\chi\pi}{2}\right)
e^{\frac{i\chi\pi}{2}}}{\sin\left(\frac{\chi\pi}{2N}\right)} =\frac{\sin\left(N\theta\right)\left[\cos\left(\theta\right)+i\sin\left(\theta\right)\right]^N}{\sin\left(\theta\right)}\nonumber\\
&=\left[\cos\left(\theta\right)+i\sin\left(\theta\right)\right]^N
   \left[\sum_{l=0}^{(N-1)/2}(-1)^lC_N^{2l+1}\cos^{N-2l-1}\theta\sin^{2l}\theta\right].
\end{eqnarray}
where $\theta=\frac{\chi\pi}{2N}$. Considering that $V_n$ must be real, i.e., $S$ must be real, namely
\begin{equation}
\sin\left(\theta\right)=0,
\end{equation}
thus $\theta=\frac{\chi\pi}{2N}=m\pi$, and $m$ should be an integer, that is, $\chi=m\times2N$, and
\begin{equation*}
\cases{
  \sin\left(m\pi\right)=0,\quad \cos(m\pi)=-1, & $m\in{\rm odd}$;\\
  \sin\left(m\pi\right)=0,\quad \cos(m\pi)=+1, & $m\in{\rm even}$.
}
\end{equation*}
Finally, we obtain
\begin{eqnarray*}
  S&= \cos^N\left(\theta\right)C_N^{1}\cos^{N-1}\theta\\
   &= N\cos^{2N-1}\theta=N(\pm1)^{2N-1}   =\cases{
      N, & $m \in \mbox{even}$;\\
      -N, & $m \in \mbox{odd}$.
    }
\end{eqnarray*}
In short,
\begin{equation}
  S=\cases{
    N, & $\pm k_1\pm k_2\pm \cdots\pm k_n=0$;\\
    N, & $\pm k_1\pm k_2\pm \cdots\pm k_n={\rm even}\times 2N$;\\
    -N, & $\pm k_1\pm k_2\pm \cdots\pm k_n={\rm odd}\times 2N$.
  }
\end{equation}
Thus we have
\begin{eqnarray}
  D_{k_1, k_2, \dots, k_n}
  &=N\left(\sum_{m=0}^{[n/2]}(-1)^{m}\sum_{\pm}\delta_{\pm k_1\pm k_2\cdots\pm k_n,\pm2N\cdot m}\right)
\end{eqnarray}
where $[\cdot]$ means round down, and $\delta$ is the Kronecker delta function. Besides, considering the symmetry of symbols ($\pm k_1$), the above formula can be further simplified as
\begin{eqnarray}
  D_{k_1, k_2, \dots, k_n}
  =2N \cdot C_{k_1, k_2, \dots, k_n}
\end{eqnarray}
where
\begin{eqnarray}
  C_{k_1, k_2, \dots, k_n}
  &=\left(\sum_{m=0}^{[n/2]}(-1)^{m}\sum_{\pm}\delta_{k_1\pm k_2\cdots\pm k_n,\pm2N\cdot m}\right).
\end{eqnarray}
Thus Eq.~(\ref{eq_AVn}) can be rewritten as
\begin{equation}
V_n=(2N)^{1-n/2} \sum_{k_1,k_2,\dots,k_n}^{N-1}Q_{k_1}Q_{k_2}\dots Q_{k_n}\omega_{k_1}\omega_{k_2}\dots \omega_{k_n}C_{k_1, k_2, \dots, k_n}.
\end{equation}

\textit{\textbf{A1. Example of $n=2$}}

\begin{eqnarray*}
  C_{k_1, k_2}
  &=\left(\sum_{m=0}^{1}(-1)^{m}\sum_{\pm}\delta_{k_1\pm k_2,\pm2N\cdot m}\right)\\
  &=\left(\sum_{\pm}\delta_{k_1\pm k_2,0}-\sum_{\pm}\delta_{k_1\pm k_2,-2N}-\sum_{\pm}\delta_{k_1\pm k_2,2N}\right).
\end{eqnarray*}
Since $1\leq k\leq N-1$, then
\begin{equation*}
  \cases{
    k_1-k_2 \in[-N+2,N-2];\\
    k_1+k_2 \in[2,2N-2].
  }
\end{equation*}
Therefore,
\begin{eqnarray*}
  C_{k_1, k_2}
  &=\delta_{k_1- k_2,0}.
\end{eqnarray*}
The coefficient $C_{k_1, k_2}$ is non-vanishing only when $k_1=k_2$. Namely, $C_{k_1, k_2}=C_{k, k}=1$.
Thus we obtain
\begin{eqnarray*}
  V_2&=\sum_{k=1}^{N-1}\omega_{k}^2Q_{k}^2.
\end{eqnarray*}

\textit{\textbf{A2. Example of $n=3$}}

\begin{eqnarray*}
  C_{k_1, k_2,k_3}
  &=\left(\sum_{m=0}^{1}(-1)^{m}\sum_{\pm}\delta_{k_1\pm k_2\pm k_3,\pm2N\cdot m}\right)\\
  &=\left(\sum_{\pm}\delta_{k_1\pm k_2\pm k_3,0}-\sum_{\pm}\delta_{k_1\pm k_2\pm k_3,-2N}-\sum_{\pm}\delta_{k_1\pm k_2\pm k_3,2N}\right).
\end{eqnarray*}
Considering that
\begin{eqnarray*}
  \cases{
    k_1-k_2-k_3 \in[-2N+3,N-3] & $\Rightarrow k_1=k_2+k_3$;\\
    k_1-k_2+k_3 \in[-N+3,2N-3] & $\Rightarrow k_1=k_2-k_3$;\\
    k_1+k_2-k_3 \in[-N+3,2N-3] & $\Rightarrow k_1=k_3-k_2$;\\
    k_1+k_2+k_3 \in[3,3N-3] &  $\Rightarrow k_1=2N-(k_2+k_3)$.\\
  }
\end{eqnarray*}
Therefore,
\begin{eqnarray*}
  C_{k_1, k_2,k_3}
  =&(\delta_{k_1- k_2-k_3,0}+\delta_{k_1- k_2+k_3,0}+\delta_{k_1+ k_2-k_3,0}-\delta_{k_1+ k_2+k_3,2N}).
\end{eqnarray*}
We obtain
\begin{eqnarray*}
  V_3&=\sqrt{\frac{1}{2N}} \sum_{k_1,k_2,k_3}^{N-1}Q_{k_1}Q_{k_2}Q_{k_3}\omega_{k_1}\omega_{k_2}\omega_{k_3}
  C_{k_1, k_2, k_3}\\
  &=\sqrt{\frac{2}{N}} \Bigg(\sum_{k_2,k_3}^{N-1}Q_{k_3-k_2}Q_{k_2}Q_{k_3}\omega_{(k_3-k_2)}\omega_{k_2}\omega_{k_3}
  \\
  &\qquad+\sum_{k_2,k_3}^{N-1}Q_{k_2+k_3}Q_{k_2}Q_{k_3}\omega_{(k_2+k_3)}\omega_{k_2}\omega_{k_3}\Bigg).
\end{eqnarray*}

In summary, the Hamiltonian (\ref{eqHam}) can be rewritten as
\begin{eqnarray}
  H = &\frac{1}{2}\sum_{k=1}^{N-1}\left(\dot{Q}_{k}^2
  +\omega_{k}^2Q_{k}^2\right)\nonumber\\
  &+\frac{\theta_n}{n(2N)^{n/2-1}} \sum_{k_1,k_2,\dots,k_n}^{N-1}Q_{k_1}Q_{k_2}\dots Q_{k_n}\omega_{k_1}\omega_{k_2}\dots \omega_{k_n}C_{k_1, k_2, \dots, k_n}.
\end{eqnarray}
The equations of motion of the normal modes can be derived as
\begin{eqnarray}
\ddot{Q}_{k}&=-\frac{\partial H}{\partial Q_k}\nonumber\\
&=-\omega_{k}^{2} Q_{k}-\frac{\theta_{n}}{(2 N)^{n / 2-1}}
\sum_{k_2,\dots,k_n}^{N-1}Q_{k_2}\dots Q_{k_n}\omega_{k}\omega_{k_2}\dots \omega_{k_n}C_{k, k_2, \dots, k_n}.
\end{eqnarray}

\section*{APPENDIX B: APPROXIMATE SOLUTION OF THE FLOQUET THEORY FOR THE FPUT-$\beta$ MODEL}

\setcounter{equation}{0}
\setcounter{figure}{0}
\makeatletter
\renewcommand{\theequation}{B\arabic{equation}}
\renewcommand{\thefigure}{B\arabic{figure}}

We now consider the expansion of the Jacobi elliptic cosine function $\rm{cn}$ in terms of trigonometric function \cite{abramowitz1965handbook}. By performing the change of variable $\tau=(\pi \Omega / 2 K(\Gamma^2))t$, Eq.~(\ref{eq_Cbeta}) can be rewritten, up to $\Gamma^{4}$ terms, as
\begin{eqnarray}\label{39}
  c(\tau) &=& -\frac{\omega_{k}^{2}}{2} \Bigg[1+\frac{3q(1+2 \cos (2 \tau ))}{2} \nonumber\\
  &~&\quad+\frac{3 q^2\left(5+16\cos (2 \tau )+4\cos (4 \tau )\right)}{32}
\Bigg],
\end{eqnarray}
where $q=\Gamma^{2}$, and $\pi$ is the period of the coefficient matrix $\boldsymbol{A}(\tau)$. Similarly, let $\boldsymbol{X}(\tau)$ is the $\left( 2\times2 \right)$ fundamental matrix that satisfies the initial condition $\boldsymbol{X}(0)=\boldsymbol{I}$. Then the system is equivalent to the following matrix equation:
\begin{equation}\label{eq_dx_tau_beta}
\frac{d}{d\tau}\boldsymbol{X}(\tau)=\boldsymbol{E}\boldsymbol{X}(\tau)+q \boldsymbol{F}(\tau)\boldsymbol{X}(\tau)+q^2 \boldsymbol{G}(\tau)\boldsymbol{X}(\tau)
\end{equation}
where
\begin{equation*}
\boldsymbol{E}=\left[\begin{array}{cc}{0} & {1} \\ {-\frac{1}{2} \omega_{k}^{2}} & {0}\end{array}\right],\quad \boldsymbol{F}(\tau)=\left[\begin{array}{cc}{0} & {0} \\ {-\frac{3\omega_{k}^{2}\left(1+2\cos(2\tau)\right)}{4}} & {0}\end{array}\right],
\end{equation*}
\begin{equation*}
\boldsymbol{G}(\tau)=\left[\begin{array}{cc}{0} & {0} \\ {-\frac{3 \omega_{k}^{2}\left(5+16\cos (2 \tau )+4\cos (4 \tau )\right)}{64}} & {0}\end{array}\right].
\end{equation*}

Following the approach presented in Ref.~\cite{leo2007stability}, the solution of Eq.~(\ref{eq_dx_tau_beta}) has the form
\begin{equation}\label{eq_Xseris}
\boldsymbol{X}(\tau)=\sum_{n=0}^\infty q^{n} \boldsymbol{X}_{n}(\tau)
\end{equation}
with all $\boldsymbol{X}_{n}(\tau)$ of class $C^{\infty}$. It is assumed that the series $\sum_{n=0}^\infty q^{n} [{X}_{n}(\tau)]_{kl}$ are all uniformly convergent with respect to $\tau$. This guarantees the derivation term to term with respect to $\tau$. Then, inserting Eq.~(\ref{eq_Xseris}) into Eq.~(\ref{eq_dx_tau_beta}), we obtain
\begin{equation}\label{eq_X0_beta}
\frac{d}{d\tau}\boldsymbol{X}_0(\tau)=\boldsymbol{E}\boldsymbol{X}_0(\tau)
\end{equation}
with $\boldsymbol{X}_0(0)=\boldsymbol{I}$;
\begin{equation}\label{eq_X1_beta}
\frac{d}{d\tau}\boldsymbol{X}_1(\tau)
=\boldsymbol{E}\boldsymbol{X}_1(\tau)
+\boldsymbol{F}(\tau)\boldsymbol{X}_{0}(\tau)
\end{equation}
with $\boldsymbol{X}_1(0)=\boldsymbol{0}$ and, for $n\geq2$, the recurrence relation
\begin{equation}\label{eq_X2_beta}
\frac{d}{d\tau}\boldsymbol{X}_n(\tau)=
\boldsymbol{E}\boldsymbol{X}_n(\tau)
+\boldsymbol{F}(\tau)\boldsymbol{X}_{n-1}(\tau)
+\boldsymbol{G}(\tau)\boldsymbol{X}_{n-2}(\tau)
\end{equation}
with $\boldsymbol{X}_n(0)=\boldsymbol{0}$.

From Eq.~(\ref{eq_X0_beta}), we have
\begin{equation}
\boldsymbol{X}_0(\tau)=e^{\boldsymbol{E}\tau}\boldsymbol{X}_0(0)=e^{\boldsymbol{E}\tau}.
\end{equation}
From Eq.~(\ref{eq_X1_beta}), we obtain
\begin{equation}
\boldsymbol{X}_1(\tau)=\int_0^\tau\boldsymbol{X}_0(\tau-s)\boldsymbol{F}(s)\boldsymbol{X}_{0}(s){\rm d}s.
\end{equation}
Similarly, for $n\geq2$, we have the recurrence form
\begin{eqnarray}
  \boldsymbol{X}_n(\tau) &=& \int_0^\tau\left[\boldsymbol{X}_0(\tau-s)\boldsymbol{F}(s)\boldsymbol{X}_{n-1}(s)+\boldsymbol{X}_0(\tau-s)\boldsymbol{G}(s)\boldsymbol{X}_{n-2}(s)\right]{\rm d}s.
\end{eqnarray}

Hereafter, we put $k=\frac{N}{2}-1$, by searching for the general solution of the approximate equation by Eq.~(\ref{eq_dx_tau_beta}), we obtain the trace of solution $\boldsymbol{X}(\pi)$, then
\begin{eqnarray}
\phi&=\frac{Tr(\boldsymbol{X}(\pi))} {2}\nonumber\\
=& \cosh \left(\pi  \sqrt{\sin \left(\frac{\pi }{N}\right)-1}\right)\left\{1+\frac{9 \pi ^2 \lambda ^2(N-1)^2}{32N^2}\left[\sin \left(\frac{\pi }{N}\right)-1\right]\right\} \nonumber\\
&\quad+\frac{3\pi  \lambda(N-1)}{64N} \sinh \left(\pi  \sqrt{\sin \left(\frac{\pi }{N}\right)-1}\right) \sqrt{\sin \left(\frac{\pi }{N}\right)-1}\nonumber\\
&\quad\times \left\{16+
\frac{\lambda(N-1)}{N} \left[12\csc \left(\frac{\pi }{N}\right)- 61\right]\right\} .
\end{eqnarray}

We know from our previous experience that $\phi<1$ (see again Fig.~\ref{figPhi}(a)), thus the instability threshold $\lambda_c$ can be obtained by solving $\phi=-1$. The complete expression of $\lambda_c$ is extremely verbose and complicated, so it is not written down here. To intuitively observe the dependence of $\lambda_c$ on the system size $N$, the form of its series expansion is given, and only the first four terms are retained, that is, Eq.~(\ref{eq_Lc_Cut_beta}) in main text.

With the help of Eq.~(\ref{eq_rho_phi}), we obtain the Floquet exponent $\mu$ as
\begin{equation}\label{eq_mu_cos_Beta}
\mu_{\pm}=\frac{\ln(\phi\pm\sqrt{\phi^2-1})}{\pi},
\end{equation}
whose explicit expression is extremely verbose, so the expression is not written out here. As an illustrating example, for fixed $N=32$, the real part of $\mu$ can be expanded as the function of $(\lambda-\lambda_c)$; i.e., Eq.~(\ref{eq_ReU2_beta_3order}) in main text.

\section*{APPENDIX C: ANALYSIS OF THE EQUATION OF ERRORS IN THE FPUT-$\alpha$ MODEL, I.E., EQ.~(\ref{26})}

\setcounter{equation}{0}
\setcounter{figure}{0}
\makeatletter
\renewcommand{\theequation}{C\arabic{equation}}
\renewcommand{\thefigure}{C\arabic{figure}}
From Eq.~(\ref{21}), we note that the coefficient does not vanish only when
\begin{equation}\label{Ceq1}
 \cases{
    k_3=k-\frac{N}{2},\quad &$k\in(\frac{N}{2},N)~~\Rightarrow~~k_3\in(0,\frac{N}{2})$;\\
    k_3=\frac{N}{2}-k,\quad &$k\in(0,\frac{N}{2})~~\Rightarrow~~k_3\in(0,\frac{N}{2})$;\\
    k_3=\frac{N}{2}+k,\quad &$k\in(0,\frac{N}{2})~~\Rightarrow~~k_3\in(\frac{N}{2},N)$;\\
    k_3=\frac{3N}{2}-k,\quad &$k\in(\frac{N}{2},N)~~\Rightarrow~~k_3\in(\frac{N}{2},N)$.\\
}
\end{equation}
Assuming $k\in(0,\frac{N}{4})$, from Eq.~(\ref{Ceq2}), we obtain two solutions of $k_3$, that is,
\begin{equation}\label{Ceq2}
  \cases{
    k_3^1=\frac{N}{2}-k\in(\frac{N}{4},\frac{N}{2});&\\
    k_3^2=\frac{N}{2}+k\in(\frac{N}{2},\frac{3N}{4}).&\\
  }
\end{equation}
Assuming $k_3^1\in(\frac{N}{4},\frac{N}{2})$, which will further excite two new modes with wave numbers:
\begin{equation}\label{Ceq3}
  \cases{
    k_3^3=\frac{N}{2}-k_3^1=k\in(0,\frac{N}{4});&\\
    k_3^4=\frac{N}{2}+k_3^1=N-k\in(\frac{3N}{4},N).&\\
  }
\end{equation}
Assuming $k_3^2\in(\frac{N}{2},\frac{3N}{4})$, which will further excite two new modes with wave numbers:
\begin{equation}\label{Ceq4}
  \cases{
    k_3^5=k_3^2-\frac{N}{2}=k\in(0,\frac{N}{4});&\\
    k_3^6=\frac{3N}{2}-k_3^2=N-k\in(\frac{3N}{4},N).&\\
  }
\end{equation}
Comparing Eq.~(\ref{Ceq3}) and Eq.~(\ref{Ceq4}), it can be seen that only the mode with $k_3^4=k_3^6$ is the newly excited. Similarly, assuming $k_3^4=N-k\in(\frac{3N}{4},N)$, it will excite two new modes with wave numbers:
\begin{equation}
  \cases{
    k_3^7=k_3^4-\frac{N}{2}=\frac{N}{2}-k\in(\frac{N}{4},\frac{N}{2});&\\
    k_3^8=\frac{3N}{2}-k_3^4=\frac{N}{2}+k\in(\frac{N}{2},\frac{3N}{4}).&\\
  }
\end{equation}
In short, assuming $k\in(0,\frac{N}{4})$, it was excited initially, we will get four modes with wave numbers of $k_3^1=k_3^7=\frac{N}{2}-k$, $k_3^2=k_3^8=\frac{N}{2}+k$, $k_3^3=k_3^5=k$, and $k_3^4=k_3^6=N-k$, which form a closed solution. In a similar way, exactly the same result is obtained when $k$ is taken in other interval. Finally, we thus obtain four-mode coupled equations~(\ref{26}).


\section*{APPENDIX D: APPROXIMATE SOLUTION OF THE FLOQUET THEORY FOR THE FPUT-$\alpha$ MODEL}

\setcounter{equation}{0}
\setcounter{figure}{0}
\makeatletter
\renewcommand{\theequation}{D\arabic{equation}}
\renewcommand{\thefigure}{D\arabic{figure}}

To study the stability of the four-mode coupled equations, let $\boldsymbol{X}(t)$ be the $\left( 8\times8 \right)$ fundamental matrix of the system of Eqs.~(\ref{26}) that satisfy the initial condition $\boldsymbol{X}(0)=\boldsymbol{I}$, where $\boldsymbol{I}$ is the $\left( 8\times8 \right)$ identity matrix. With the aid of $Q_{\frac{N}{2}}=\sqrt{\varepsilon(N-1)} \cos \left(\tau\right)$, where $\tau=\sqrt{2}t$, the system is equivalent to the following matrix equation:
\begin{equation}\label{28}
\frac{d}{d\tau}\boldsymbol{X}(\tau)=\boldsymbol{K}\boldsymbol{X}(\tau)+\gamma \boldsymbol{L}(\tau)\boldsymbol{X}(\tau)
\end{equation}
where $\gamma=2\alpha \sqrt{\varepsilon(N-1)/N}=2\sqrt{\lambda(N-1)/N}$, and
\begin{equation*}
\boldsymbol{K}=\left[\begin{array}{cccccccc}{0} & {1} & {0} & {0} & {0} & {0} & {0} & {0} \\ {-\omega_{k}^{2}} & {0} & {0} & {0} & {0} & {0} & {0} & {0} \\ {0} & {0} & {0} & {1} & {0} & {0} & {0} & {0} \\ {0} & {0} & {-\omega_{\frac{N}{2}-k}^{2}} & {0} & {0} & {0} & {0} & {0} \\ {0} & {0} & {0} & {0} & {0} & {1} & {0} & {0} \\ {0} & {0} & {0} & {0} & {-\omega_{\frac{N}{2}+k}^{2}} & {0} & {0} & {0} \\ {0} & {0} & {0} & {0} & {0} & {0} & {0} & {1} \\ {0} & {0} & {0} & {0} & {0} & {0} & {-\omega_{N-k}^{2}} & {0}\end{array}\right],
\end{equation*}
and $\boldsymbol{L}(\tau)$ is also a matrix whose elements that differ from zero are
\begin{eqnarray*}
l_{23}=l_{41}&=-\omega_{k}\omega_{\frac{N}{2}-k} \cos(\tau),\quad
l_{25}=l_{61}=-\omega_{k}\omega_{\frac{N}{2}+k} \cos(\tau),\\
l_{47}=l_{83}&=-\omega_{\frac{N}{2}-k}\omega_{N-k} \cos(\tau),\quad
l_{67}=l_{85}=\omega_{\frac{N}{2}+k}\omega_{N-k} \cos(\tau),
\end{eqnarray*}
respectively.

In a similar way, inserting Eq.~(\ref{eq_Xseris}) into Eq.~(\ref{28}), we obtain
\begin{equation}\label{eq0}
\frac{d}{d\tau}\boldsymbol{X}_0(\tau)=\boldsymbol{K}\boldsymbol{X}_0(\tau)
\end{equation}
with $\boldsymbol{X}_0(0)=\boldsymbol{I}$ and, for $n\geq1$, the recurrence relation
\begin{equation}\label{eq1}
\frac{d}{d\tau}\boldsymbol{X}_n(\tau)=\boldsymbol{K}\boldsymbol{X}_n(\tau)+\boldsymbol{L}(\tau)\boldsymbol{X}_{n-1}(\tau)
\end{equation}
with $\boldsymbol{X}_n(0)=\boldsymbol{0}$. From Eq.~(\ref{eq0}) and Eq.~(\ref{eq1}), one gets
\begin{equation}\label{eq2}
\boldsymbol{X}_0(\tau)=e^{\boldsymbol{K}\tau}\boldsymbol{X}_0(0)=e^{\boldsymbol{K}\tau},
\end{equation}
and the recurrence relation
\begin{equation}\label{eq2b}
\boldsymbol{X}_n(\tau)=\int_0^\tau\boldsymbol{X}_0(\tau-s)\boldsymbol{L}(s)\boldsymbol{X}_{n-1}(s){\rm d}s.
\end{equation}
In particular, for $\tau=2\pi$ , the period of matrix $\boldsymbol{L}(\tau)$,
the characteristic numbers $\rho$ of the system described by Eq.~(\ref{28}), which are also eigenvalues of matrix $\boldsymbol{X}(2\pi)$. If $|\rho_j|<1$ for $j\in[1,8]$, then the system is stable. On the contrary, as long as one of them exceeds $1$, the system is unstable.

In principle, the stability analysis of the system of Eq.~(\ref{28}) should be made for each $k\in(0,N/4)$, which determines the elements of $\boldsymbol{K}$ and $\boldsymbol{L}(\tau)$. However, from our numerical results (see Fig.~\ref{figEkT}(b)), we clearly see that the adjacent modes; i.e., $k=N/2\pm1$ become unstable first. Hence, the instability threshold of the system will be estimated by setting $k=1$, such that $\boldsymbol{K}$ and $\boldsymbol{L}(\tau)$ are settled.

In practice, Eq.~(\ref{eq_Xseris}) has to be truncated. Here the equation is only retained to the term of $\gamma^2$. Namely,
\begin{equation}\label{32b}
 \boldsymbol{X}(2\pi)=\boldsymbol{X}_{0}(2\pi)+\gamma\boldsymbol{X}_{1}(2\pi)+\gamma^2\boldsymbol{X}_{2}(2\pi)+O\left(\gamma^3\right),
\end{equation}
by assuming $0<\gamma\ll1$. The iterative operation of $\left( 8\times8 \right)$ matrix is much more complicated, it has been performed with the aid of the MATLAB. Then using the method in Ref.~\cite{chechin2012stability}, finally, we obtain Eq.~(\ref{eq_Lc_alpha}) in main text.

\section*{References}

\bibliography{ref}

\end{document}